\UseRawInputEncoding

\documentclass[letterpaper]{cas-dc}

\usepackage[numbers]{natbib}

\usepackage{amssymb,mathrsfs,bm} 
\usepackage{graphicx}

\usepackage[labelformat=simple]{subcaption}

\usepackage{enumitem}
\usepackage{float}

\usepackage{dcolumn}
\usepackage{siunitx}
\usepackage{pifont}
\usepackage{mathtools}

\usepackage{multirow}

\DeclareMathOperator{\sinc}{sinc}

\newcommand{\rd}{\mathrm{d}}
\newcommand{\re}{\mathrm{e}}
\newcommand{\Da}{\mathcal{D}}

\usepackage{hyperref}
\hypersetup{
    colorlinks=true,
    linkcolor=blue,
    filecolor=magenta,
    urlcolor=cyan,
    citecolor=blue,
    breaklinks=true
}

\allowdisplaybreaks

\begin{document}
\let\WriteBookmarks\relax
\def\floatpagepagefraction{1}
\def\textpagefraction{.001}

\shorttitle{Non-Fickian macroscopic model of axial diffusion of granular materials}

\shortauthors{I C Christov and H A Stone}

\title[mode = title]{Non-Fickian macroscopic model of axial diffusion of granular materials in a long cylindrical tumbler}


\author[1,2]{Ivan~C.\ Christov}[orcid=0000-0001-8531-0531]
\cormark[1]
\fnmark[1]
\ead{christov@purdue.edu}
\credit{Conceptualization, Methodology, Investigation, Formal analysis, Writing - Original Draft}
\address[1]{School of Mechanical Engineering, Purdue University, West Lafayette, Indiana 47907, USA}
\address[3]{Department of Computer Science, University of Nicosia, 46 Makedonitissas Avenue, CY-2417, Nicosia, Cyprus}

\cortext[cor1]{Corresponding author}
\fntext[fn1]{Also affiliated with the Center for Particulate Products and Processes (CP\textsuperscript{3}) at Purdue University.}

\author[4]{Howard~A.\ Stone}[orcid=0000-0002-9670-0639]
\ead{hastone@princeton.edu}
\credit{Conceptualization, Methodology, Writing - Original Draft}
\address[4]{Department of Mechanical and Aerospace Engineering, Princeton University, Princeton, New Jersey 08544, USA}

\begin{abstract}
We provide new concepts for understanding transport phenomena in flows of granular materials by using a non-Fickian macroscopic model of axial diffusion of a granular material in a finite cylindrical tumbler. The model accounts for diffusion induced by particle collisions only in a thin surface flowing layer due to localization of shear within the cross-section of the drum. All model parameters are related to measurable quantities in a granular flow. It is shown that the proposed model is a member of the general class of linear constitutive relations with memory. An exact solution for the spreading of a finite-width pulse initial condition under the proposed non-Fickian model is derived and compared to the solution of a Fickian model (i.e., the ``classical'' diffusion equation).
\end{abstract}

%


\maketitle

\section{Introduction}
\label{sec:intro}

Granular materials at rest, in flow or under external agitation behave like various continua (liquid, solid, gases), albeit with quite ``unusual'' properties \cite{jnb96,at09}. The continuum approach has allowed successful practical modeling of granular mixing and segregation \cite{ok00,mlo07,ulo19}. Yet, ``[d]espite an expanding body of literature in the last ten to fifteen years, relationships between identified mechanisms are ambiguous, experimental data is scarce, and there is no accepted model[s],'' as Rosato and Blackmore note in the preface to the 1999 IUTAM Symposium on Segregation in Granular Flows \cite{rb00}. Arguably, this statement remains true today, even after two more decades of progress in granular mechanics. 

\begin{figure}
\centering
  \includegraphics[width=\columnwidth]{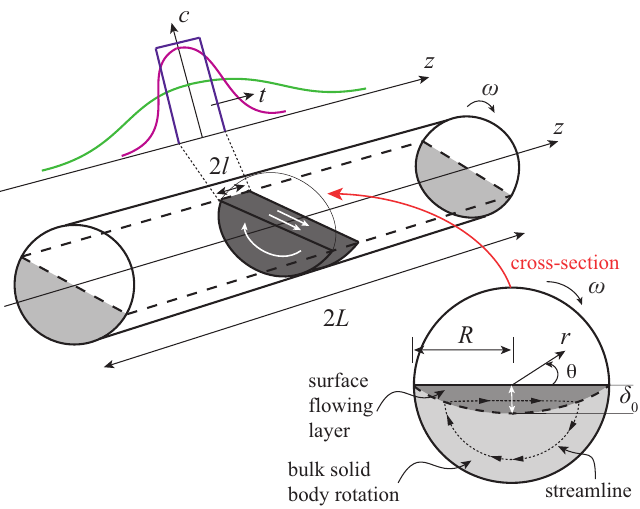}
\caption{Schematic showing the flow geometry and notation for a long cylindrical tumbler, of radius $R$ and length $2L$, rotating with angular frequency $\omega$; adapted from \cite{cs12}. The bottom-right inset is a diagram of the flow in the tumbler's cross-section, in which the thin surface flowing shear layer has maximal depth $\delta_0$. The top-left overlay is a diagram of the cross-sectionally averaged axial diffusion process of a band, of initial width $2l$, of particles whose cross-sectionally averaged concentration is $c(z,t)$. Shaded regions are the filled portions of the tumbler (a fill fraction $\phi=0.5$ is illustrated).}
\label{fig:tumbler}
\end{figure}

Diffusion is a fundamental powder mixing (transport) mechanism modeled using continuum theories. A number of agitation processes lead to non-equilibrium velocity fluctuations that cause ``self-diffusion'' in granular flows: from tumblers \cite{zs91} and vertically vibrated beds \cite{rlr08} (axial diffusion) to ``simple'' shear flows (anisotropic diffusion) \cite{s93,nht95,c97} to horizontal \cite{bl89} and vertical \cite{hk05} Hele-Shaw cells (radial diffusion) to cylindrical Couette cells (axial and radial diffusion) \cite{ub04,jb04}. Then, axial mixing of granular materials in a rotating drum, as shown schematically in Fig.~\ref{fig:tumbler}, is typically modeled using Fick's second law, leading to the ``classical'' diffusion equation \cite{l54,cfhhr66,hchf66,dr99}. For example, Cahn et al.~\cite{cfhhr66} provided early experimental justification for a ``diffusive'' axial transport process by tracking the spread of yellow beads into identical white beads, of diameter $d_p=200$ $\mu$m, in a cylindrical drum of radius $R=0.1$ m (note that $d_p \ll R$ is required for this continuum description to hold).

Granular flow in a long drum has proven to be a simple but important system to understand \cite{st11}. Experiments  in the 2000s \cite{km05,ffas09} reinvigorated interest in this flow because of the possibility of \emph{anomalous diffusion}. Although ``anomalous'' \emph{scalings} can also be interpreted within the theory of intermediate self-similar asymptotics of ``classical'' diffusion \cite{cs12}, the fundamentals of granular diffusion are still poorly understood. Macroscopic models based on Fick's laws are not derived from first principles but rather fit to experimental or particle dynamics (i.e., simulation) data \cite{tss10,mtm15}. Nevertheless, experiments have established the key features of granular flow in rotating containers \cite{CaMD2,bb97,kmso97,jol02}. The most important conclusion drawn is that the flow in an axial cross-section of a long rotating drum (Fig.~\ref{fig:tumbler}) is composed of a thin surface shear layer (the \emph{flowing layer}) in which material flows quickly down the slope. Particles below it (the \emph{fixed bed}) are in static equilibrium performing solid body rotation in unison with the container. Acknowledging these disparate flow characteristics in the cross-section, Das Gupta et al.~\cite{dgkb91} related the axial diffusivity and drift velocities in a tilted tumbler to its rotation rate, tilt angle, and the particle properties. Their model starts from the stochastic equations for a single particle and necessitates the prescription of a probability distribution of axial displacements due to interparticle collisions. 

As Metcalfe and Shattuck note, ``[i]t takes more than 200 revolutions to disperse the seeds ... axially along the tube boundary and then mix radially. This is physically plausible given that the only motions mixing the material are the avalanches across the free surface'' \cite{ms96}. Thus, if the surface flow in the tumbler is the only mechanism that leads to transport and mixing (as also emphasized in Fig.~1(c,d) in \cite{cfhhr66}), then the axial spread of particles should be expected to be more complicated than the Fickian diffusion models originally proposed in \cite{hchf66,dgkb91}. Specifically, the fluctuations in the axial direction are distinct from the random displacements that lead to radial mixing in the cross-section. Given that the latter dynamics are rapidly equilibrated \cite{kmso97,st11}, the radial transport mechanisms in the tumbler are not of interest in this work. Instead, here, we would like to address a basic scientific question about the axial transport: \emph{Does taking into account the disparate flow characteristics in the tumbler cross-section lead to a non-Fickian (but not necessarily ``anomalous'') axial diffusion equation?} Our goal is to provide new theoretical insights into transport phenomena in flows of granular materials.

\section{A macroscopic axial diffusion model accounting for localization of shear in the cross-section}
\label{sec:model}

Let $c(z,t) = \frac{1}{\phi A}\iint_A \hat{c}(r,\theta,z,t) \,\rd A$ be the cross-sec\-tio\-nal\-ly averaged concentration in a partially filled cylinder with cross-sectional area $A$ and fill fraction $\phi$;  $\hat{c}$ is the number of particles per unit volume. In the cross-section of the tumbler, particles continuously exchange between the surface flowing layer (in which collisions lead to random axial displacement and, thus, diffusion) and a fixed bed in which particles only rotate with the tumbler and cannot be displaced axially. {The granular material is assumed to be monodisperse.} Thus, to study the axial transport process, we label a certain proportion of the particles as ``diffusing,'' $c_\mathrm{d}$, and the remainder as ``non-diffusing,'' $c_\mathrm{nd}$; $c = c_\mathrm{d} + c_\mathrm{nd}$. Within the cross-sectionally averaged description, we envision that the diffusing ``species'' concentration, $c_\mathrm{d}$, can represent tagged particles being observed to spread axially. Consequently, in any axial cross-section along the tumbler, there can be both diffusing and non-diffusing particles, and their concentrations will vary in space, $z$, and time, $t$. Now, we wish to develop a transport model for the spatiotemporal evolution of $c_\mathrm{d}$ and $c_\mathrm{nd}$, accounting for the flow features thus described.

A ``minimal'' model of such a diffusion process (see, e.g., \cite[Sec.~3.3.1]{mcb14}) is%
\begin{subequations}\begin{align}
\frac{\partial c_\mathrm{d}}{\partial t} &= D\frac{\partial^2 c_\mathrm{d}}{\partial z^2} - k(c_\mathrm{d} - \beta c_\mathrm{nd}),\label{eq:cd}\\
\frac{\partial c_\mathrm{nd}}{\partial t} &= k(c_\mathrm{d} - \beta c_\mathrm{nd}),\label{eq:cnd}
\end{align}\label{eq:react_diff}\end{subequations}
subject to given initial conditions (ICs) on the species:
\begin{equation}
c_\mathrm{d}(z,0) = c_{\mathrm{d},0}(z),\qquad c_\mathrm{nd}(z,0) = c_{\mathrm{nd},0}(z),
\label{eq:rd_ic}
\end{equation}
and no-flux boundary conditions (BCs) at the tumbler's endwalls:
\begin{equation}
\left.\frac{\partial c_\mathrm{d}}{\partial z}\right|_{z=\pm L} = \left.\frac{\partial c_\mathrm{nd}}{\partial z}\right|_{z=\pm L} = 0.
\label{eq:rd_bc}
\end{equation}

In Eq.~\eqref{eq:cd}, $D$ is an ``effective'' diffusivity [dimensions of (length)\textsuperscript{2}/time] characterizing the mean-squared axial displacements caused by particle collisions in the flowing layer; $k$ is a rate constant (dimensions of 1/time) characterizing the flowing layer--fixed bed particle exchange using first-order kinetics; and, the dimensionless parameter $\beta$, which is $>0$ and $\ne 1$, is a \emph{partition} ratio (or coefficient) related to the fact that the flowing layer and the fixed bed take up unequal areas of the cross-section. Another way to interpret $\beta$ from the mathematical structure of the model, is that there is lack of \emph{local} conservation (in a given axial cross-section) between the two species, due to the fact that the diffusing species can leave the given cross-section (while the non-diffusing species cannot), which leads to unequal exchange rates (i.e., $k\ne\beta k$) between diffusing the non-diffusing species. Nevertheless, a \emph{global} conservation law follows from summing Eqs.~\eqref{eq:react_diff}, integrating over space, and using the BCs from Eq.~\eqref{eq:rd_bc}:%
\begin{equation}
\frac{\rd}{\rd t} \int_{-L}^{+L} \big( c_\mathrm{d} + c_\mathrm{nd} \big) \,\mathrm{d}z = \left. D\frac{\partial c_\mathrm{d}}{\partial z}\right|_{-L}^{+L} = 0.
\end{equation}

More generally, Eqs.~\eqref{eq:react_diff} represent a reaction-diffusion system in which there is \emph{local immobilization} of one species \cite{rs13}. Such equations also arise in the modeling of signaling and transport in certain biological systems \cite{bcs09,detal10} wherein proteins can become ``mobile'' (diffusing) or ``immobile'' (non-diffusing) as they bind or unbind from a local substrate. A similar situation occurs in electrodeposition \cite{lab99,bs00}, during flows in fractured reservoirs \cite{bzk60} and dusty gasses \cite{s62,l66}, and during heat transfer between an electron gas and a metal lattice \cite{toc94}, among other examples (see also \cite{c08}).

Equations~\eqref{eq:react_diff} can be written as a single partial differential equation (PDE) for $c_\mathrm{d}$ by taking $\partial/\partial t$ of Eq.~\eqref{eq:cd}, eliminating $\partial c_\mathrm{nd}/\partial t$ using Eq.~\eqref{eq:cnd}, then eliminating $k(c_\mathrm{d} - \beta c_\mathrm{nd})$ using Eq.~\eqref{eq:cd}:
\begin{equation}
k(1+\beta) \frac{\partial c_\mathrm{d}}{\partial t} + \frac{\partial^2 c_\mathrm{d}}{\partial t^2} = D\left (k\beta\frac{\partial^2 c_\mathrm{d}}{\partial z^2} + \frac{\partial^3 c_\mathrm{d}}{\partial t\partial z^2}\right).
\label{eq:cd_single_eq}
\end{equation}
Observe that Eq.~\eqref{eq:cd_single_eq} features both second-order-in-time and mixed derivatives, thus some authors designate it as a dissipative \emph{wave} equation \cite{c08}. Importantly, Eq.~\eqref{eq:cd_single_eq} \emph{cannot} be derived from Fick's first and second law, thus it is termed \emph{non-Fickian}.

Next, we determine how the parameters $k$ and $\beta$ in Eqs.~\eqref{eq:react_diff} relate to the physics of axial diffusion of a granular material in a tumbler (Fig.~\ref{fig:tumbler}).

\paragraph{Determining $\beta$:}
At steady-state ($t\to\infty$) we expect both spatial and temporal gradients to vanish {for a monodisperse granular material. (Spatial gradients in concentration may exist in size-bidisperse systems even at steady state; see, e.g., \cite{chol10}.)} Then, we may seek \emph{constant} solutions to Eq.~\eqref{eq:react_diff} and Eq.~\eqref{eq:rd_bc}, and we find that
\begin{subequations}\begin{align}
\lim_{t\to\infty}c_\mathrm{d}(z,t) &= \frac{\beta}{1+\beta}(c_{\mathrm{d},0} + c_{\mathrm{nd},0}) = \frac{\beta}{1+\beta},\label{eq:cd_ss}\\
\lim_{t\to\infty}c_\mathrm{nd}(z,t) &= \frac{1}{1+\beta}(c_{\mathrm{d},0} + c_{\mathrm{nd},0}) = \frac{1}{1+\beta},
\end{align}\label{eq:cd_cnd_ss}\end{subequations}
where $c_{\mathrm{d},0}$ and $c_{\mathrm{nd},0}$ must now be constants given the assumption of no axial gradients. Note that $c_{\mathrm{d},0} + c_{\mathrm{nd},0} = 1$ can always be enforced due to the linearity of Eqs.~\eqref{eq:react_diff}.

The cross-section of the cylinder in Fig.~\ref{fig:tumbler} is circular, hence the area occupied by particles is $A=\phi\pi R^2$. To a good approximation, the flowing layer's shape is an ellipse with semi-major axis $R$ and semi-minor axis $\delta_0$ \cite[\S3]{mlo07}. The maximal depth of the flowing layer $\delta_0$ is typically on the order of $0.1R$ to $0.2R$ \cite{kmso97} (or, about 8 to 12 particles thick for beads a few mm in diameter in a $24$ cm-diameter drum \cite{jol02}). Thus, the flowing layer's fraction of the cross-sectional area is
\begin{equation}
\frac{\tfrac{1}{2}\pi \delta_0 R}{\phi \pi R^2} = \frac{1}{2\phi}\left(\frac{\delta_0}{R}\right).
\label{eq:fl_frac}
\end{equation}
At steady state, the fractions of particles in the flowing layer, as given by Eq.~\eqref{eq:cd_ss} and Eq.~\eqref{eq:fl_frac}, must match, thus%
\begin{equation}
\beta = \frac{\delta_0}{2\phi R-\delta_0} \simeq \frac{1}{2\phi}\left(\frac{\delta_0}{R}\right) \qquad (\delta_0\ll R).
\label{eq:beta}
\end{equation}

\paragraph{Determining $k$:}
The constant $k$ quantifies the rate at which the diffusing species leaves the flowing layer. The axial diffusion process under consideration occurs on a longer time scale compared to the radial transport. This separation of time scales is supported by experiments, which have shown that radial transport equilibrium is rapidly established in a drum, and particles are exchanged between the flowing layer and the bulk in a time-periodic, or quasi-steady, manner \cite{kmso97,st11}. Now, the flux through the flowing layer's interface with the bulk can be estimated as the area of the flowing layer, $\tfrac{1}{2}\pi\delta_0 R$, times the rate-of-loss (or gain) of the diffusing species, $k$, or $Q_\mathrm{d} \simeq \tfrac{1}{2}\pi\delta_0 R k$. 

At the same time, for partially filled tumblers ($\phi \lesssim 50\%$), all particles complete a transit of the cross-section in about half a revolution, specifically also entering (and leaving) the flowing layer in that time \cite{ok00}. So, the flux can also be estimated as the total filled area, $A=\phi\pi R^2$, divided by the half period of rotation, $\pi/\omega$, yielding $Q \simeq \tfrac{1}{2}\omega \phi R^2$. For quasi-steady flow in the cross-section, we must have $Q=Q_\mathrm{d}$, hence we find that%
\begin{equation}
k\simeq \frac{\omega}{\pi}\phi\left(\frac{\delta_0}{R}\right)^{-1}.
\label{eq:k}
\end{equation}
Again, typically $\delta_0\ll R$, so $k/\omega\gg 1$.

\paragraph{Nondimensionalization:} Consistent with prior literature, we use the tumbler's axial half-length as the length scale and the axial diffusion time as the time scale: $z = LZ$, $t = (L^2/D)T$. Then, letting $C_{\mathrm{d}}(Z,T) = c_{\mathrm{d}}(z,t)$, Eq.~\eqref{eq:cd_single_eq} becomes
\begin{equation}
(1+\beta) \frac{\partial C_\mathrm{d}}{\partial T} + \Da \frac{\partial^2 C_\mathrm{d}}{\partial T^2} = \beta\frac{\partial^2 C_\mathrm{d}}{\partial Z^2} + \Da \frac{\partial^3 C_\mathrm{d}}{\partial T\partial Z^2},
\label{eq:cd_single_eq_nd1}
\end{equation}
Here, $\Da = D/(kL^2) =  (\pi \delta_0 D)/(\omega R L^2)$ is a dimensionless parameter---a type of Damk\"ohler number---that represents the ratio of the ``reaction time scale'' ($1/k$) to the diffusion time scale ($L^2/D$). We expect that $\Da \ll 1$ since the reaction time scale is related to the exchange of particles between the flowing layer and the bulk, which occurs continuously as the tumbler rotates, while diffusion in the axial direction occurs more slowly over many revolutions. As an example, for particles with diffusivity on the order of $D\approx1$ mm\textsuperscript{2}/s \cite{dr99,soc04,tss10} in a thin flowing layer with aspect ratio $\delta_0/R\approx0.1$ in a tumbler of length $2L=600$ mm and radius $R=14.25$ mm rotating at $\omega = 0.62$ rev/s $=1.24\pi$ rad/s \cite{km05}, we obtain the rough estimate $\Da \simeq 10^{-6}$.

Notice that, for $\Da \to 0^+$ (e.g., $k\to\infty$, an infinitely fast flowing layer--fixed bed exchange),  Eq.~\eqref{eq:cd_single_eq_nd1} becomes the Fickian ``classical'' diffusion equation postulated in previous works \cite{l54,hchf66,dgkb91} \emph{but} the diffusivity is multiplied by $\beta/(1+\beta)\sim\delta_0/R$ (for $\delta_0\ll R$). Therefore, when fitting solutions of the Fickian diffusion equation to data, the diffusivity may be overestimated by at least an order of magnitude, depending on the conditions under which the data is taken.

Next, we solve the axial granular diffusion initial-bound\-ary value problem (IBVP) for Eq.~\eqref{eq:cd_single_eq_nd1}. Specifically, we seek to highlight the effect of $\Da > 0$ on the transient axial diffusion process. As argued in \cite{cs12}, the time required for the concentration profile to relax away from a finite-width pulse IC can be one reason for observing ``anomalous'' scalings.

\section{The initial-boundary value problem on $Z\in[-1,+1]$}
\label{sec:ibvp}

First, we must supplement Eq.~\eqref{eq:cd_single_eq_nd1} with appropriate ICs, say $C_\mathrm{d}(z,0) = C_{\mathrm{d},0}(Z)$ and $(\partial C_\mathrm{d}/\partial T)_{T=0} = C_{\mathrm{d},1}(Z)$. $C_{\mathrm{d},1}(Z)$ is not arbitrary because it must satisfy a compatibility condition based on Eqs.~\eqref{eq:react_diff}. Specifically, from Eq.~\eqref{eq:cd}, $C_{\mathrm{d},1}(Z)$ can be related to the non-diffusing species' initial condition $C_{\mathrm{nd},0}(Z)$ [recall Eq.~\eqref{eq:rd_ic}] as $C_{\mathrm{d},1}(Z) = \Da \rd^2 C_{\mathrm{d},0}(Z)/\rd Z^2 - C_{\mathrm{d},0}(Z) + \beta C_{\mathrm{nd},0}(Z)$. Now, for the problem at hand, suppose an equilibrium partition of diffusing and non-diffusing species in the cross-section at $T=0$, in which case $\beta {C}_\mathrm{nd}(Z,0) = {C}_\mathrm{d}(Z,0)$. Then, a band of unit area of diffusing particles is tagged, and we seek to determine its evolution. To this end, normalize $C_\mathrm{d}$ as $\tilde{C}_\mathrm{d}(Z,T) = C_\mathrm{d}(Z,T)/C_{\mathrm{d},\infty}$ where $C_{\mathrm{d},\infty}=\beta/(1+\beta)$ is the constant steady-state distribution from Eq.~\eqref{eq:cd_ss}. The ICs for Eq.~\eqref{eq:cd_single_eq_nd1} then become%
\begin{subequations}\begin{align}
\tilde{C}_\mathrm{d}(Z,0) &= \frac{1}{2\ell}\left[H(Z+\ell) - H(Z-\ell)\right],\\
\frac{\partial\tilde{C}_\mathrm{d}}{\partial T}(Z,0) &= \frac{\Da}{2\ell}\left[\delta'(Z+\ell) - \delta'(Z-\ell)\right],
\end{align}\label{eq:pulse_ics}\end{subequations}
where $2\ell < 2$ is the band's width ($\ell = l/L$ is dimensionless), $H(\cdot)$ is the Heaviside unit step function, $\delta(\cdot)$ is the Dirac-delta distribution and $\delta' \equiv \rd \delta/\rd Z$. The BCs given in Eq.~\eqref{eq:rd_bc} carry over to $\tilde{C}_\mathrm{d}$.

Next by separation of variables using the eigenfunctions of $\rd^2/\rd Z^2$ satisfying homogeneous Neumann BCs, the solution can be written as a Fourier cosine series: $\tilde{C}_\mathrm{d}(Z,T) = \tfrac{1}{2}a_0(T) + \sum_{n=1}^\infty a_n(T)\cos(n\pi Z)$. Substituting into Eq.~\eqref{eq:cd_single_eq_nd1} and employing orthogonality of the eigenfunctions:%
\begin{equation}
\Da\frac{\rd^2 a_n}{\rd T^2} + \left[ (1+\beta) + n^2\pi^2 \Da\right] \frac{\rd a_n}{\rd T} + n^2\pi^2 \beta a_n = 0.
\label{eq:coeff_ode}
\end{equation}
The ICs from Eqs.~\eqref{eq:pulse_ics} give
\begin{subequations}\begin{align}
a_n(0) &= \sinc(n\pi\ell),\\
\left.\frac{\rd a_n}{\rd T}\right|_{T=0} &= -n^2\pi^2\Da a_n(0),
\end{align}\end{subequations}
for $n\ge0$. Here, we have defined $\sinc \eta = \eta^{-1}\sin\eta$ for real $\eta\ne0$ and $\sinc(0)=1$. Applying the temporal Laplace transform, $\overline{(\cdot)} = \int_0^\infty (\cdot) \,\re^{-sT} \,\rd T$ ($s\in\mathbb{C}$), to Eq.~\eqref{eq:coeff_ode}, we obtain, for $n\ge0$,
\begin{equation}
\overline{a_n}(s) = \frac{a_n(0) [s + (1+\beta)/\Da + n^2\pi^2(1-\Da)]}{\left\{s^2 + s[(1+\beta)/\Da + n^2\pi^2 ]+ n^2\pi^2 \beta/\Da\right\}}.
\label{eq:an_bar_s}
\end{equation}

Following \cite{c13}, Eq.~\eqref{eq:an_bar_s} can be inverted back to the $t$ domain by partial fractions and tables of Laplace transforms, to yield $a_0(t) = H(t)$, and%
\begin{multline}
a_n(T) = \sinc(n\pi\ell) H(T) \exp\left(-\Upsilon_n T\right) 
\left[ \cosh\left(\displaystyle\frac{T\sqrt{\Delta_n}}{2\Da}\right) \right. \\  \left. + \frac{(1+\beta)+n^2\pi^2\Da(1-2\Da)}{\sqrt{\Delta_n}}
\sinh\left(\displaystyle\frac{T\sqrt{\Delta_n}}{2\Da}\right) \right], 
\label{eq:an_t}
\end{multline}
for $n>0$, where we have defined $\Delta_n:=\Delta(n\pi)$ and $\Upsilon_n:=\Upsilon(n\pi)$, with 
\begin{subequations}\begin{align}
\Upsilon(\xi) &:= (1+\beta+\xi^2\Da)/(2\Da),\\
\Delta(\xi) &:= (1+\beta+\xi^2\Da)^2 - 4\xi^2 \beta\Da,
\end{align}\end{subequations}
for convenience. In general, we must consider the three cases $\Delta_n \gtreqqless 0$. On physical grounds, we expect $\beta < 1$ (indeed, $\beta\ll 1$) as discussed in Sec.~\ref{sec:model}. Then, it can be shown $\Delta(\xi) > 0$ $\forall \xi$ and $\Da$, thus we may take $\Delta_n>0$. 

To summarize, the \emph{exact} dimensionless solution to the IBVP comprised of Eq.~\eqref{eq:cd_single_eq_nd1}, Eq.~\eqref{eq:pulse_ics} and Eq.~\eqref{eq:rd_bc} is 
\begin{equation}
C_\mathrm{d}(Z,T) =  \frac{\beta}{1+\beta}\left[ \frac{1}{2}H(T) + \sum_{n=1}^\infty a_n(T) \cos(n\pi Z) \right],
\label{eq:cd_exact_soln_pulse}
\end{equation}
where $a_n(T)$ is given in Eq.~\eqref{eq:an_t}. To the best of our knowledge, Eq.~\eqref{eq:cd_exact_soln_pulse} is a new exact solution [in any context in which Eq.~\eqref{eq:cd_single_eq_nd1} arises]. The corresponding IVP on $Z\in(-\infty,+\infty)$ is discussed in \cite[Sec.~V-A]{rs13}, including the derivation of the fundamental solution [i.e., the solution to the IVP with ICs $C_\mathrm{d}(Z,0) = \delta(Z)$ and $(\partial C_\mathrm{d}/\partial T)_{T=0} = 0$]. The fundamental solution was found to consist of a singular part localized at the origin (but decaying exponentially in time) and a regular part that behaves asymptotically as the well-known Gaussian solution of the diffusion equation [i.e., Eq.~\eqref{eq:cd_single_eq_nd1} with $\Da=0$]. Unlike the latter case, all terms in Eq.~\eqref{eq:cd_exact_soln_pulse} are regular (i.e., not singular). With $C_\mathrm{d}(Z,T)$ now determined, $C_\mathrm{nd}(Z,T)$ can be easily found from the dimensionless version of Eq.~\eqref{eq:cnd}.

Note, however, that the term premultiplying $\sinh(\cdot)$ in Eq.~\eqref{eq:an_t} scales as $n$, which is expected to cause a significant Gibbs phenomenon in the final Fourier series solution in Eq.~\eqref{eq:cd_exact_soln_pulse}, when $\Da \not\to 0$. As suggested in \cite{kjc18}, to mitigate the Gibbs phenomenon when making plots, the Fourier series from Eq.~\eqref{eq:cd_exact_soln_pulse} is evaluated by multiplying each term in the summation over $n$ by a cubic Lanczos $\sigma$-factor~\cite[pp.~221--227]{L56}, i.e., $\sinc^3(n\pi/N)$, where $n=N-1$ is the final term in the truncated series. The $\sigma$-factors reduce the Gibbs phenomenon without affecting the convergence of the Fourier series.

As $T\to\infty$, the non-Fickian solution given by Eq.~\eqref{eq:cd_exact_soln_pulse} is expected to converge to the equivalent solution of the Fickian diffusion equation [i.e., Eq.~\eqref{eq:cd_single_eq_nd1} with $\Da=0$] \cite{rs13}, which can be computed following the same steps as above:
\begin{multline}
C_\mathrm{d}(Z,T;\Da=0) = \frac{\beta}{1+\beta}H(T) \\ \times \left[ \frac{1}{2} + \sum_{n=1}^\infty  \re^{-n^2\pi^2 D_\mathrm{F}\, T} \sinc(n\pi\ell) \cos(n\pi Z) \right],
\label{eq:cd_exact_soln_pulse_Da0}
\end{multline}
where $D_\mathrm{F} := \beta/(1+\beta)$ is dimensionless.

\begin{figure*}
\centering
\includegraphics[width=\textwidth]{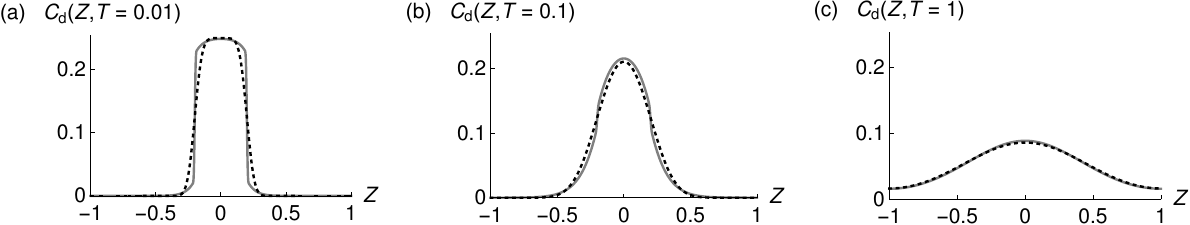}
\vskip 2mm
\includegraphics[width=\textwidth]{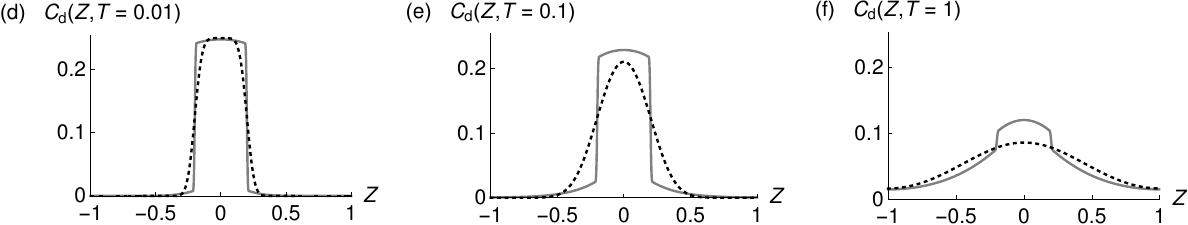}
\caption{Time-evolution snapshots of the (dimensionless) diffusing species concentration $C_\mathrm{d}$ from the non-Fickian solution in Eq.~\eqref{eq:cd_exact_soln_pulse} as solid curves, at (a,d) $T=0.01$, (b,e) $T=0.1$, (c,f) $T=1$ for (a,b,c) a Damk\"ohler-like number $\Da = 5\times10^{-3}$ and (d,e,f) $\Da = 5\times10^{-2}$. The partition ratio is $\beta = 1/9$ and the half-width of the pulse is $\ell = 1/5$. Dashed curves correspond to the Fickian diffusion solution ($\Da=0$) given in Eq.~\eqref{eq:cd_exact_soln_pulse_Da0}.}
\label{fig:results1}
\end{figure*}

Figure~\ref{fig:results1} shows, respectively for $\Da=5\times10^{-3}$ (a,b,c) and $\Da=5\times10^{-2}$ (d,e,f), comparisons between the exact solutions given by Eq.~\eqref{eq:cd_exact_soln_pulse} (solid) and Eq.~\eqref{eq:cd_exact_soln_pulse_Da0} (dashed) at different $T$. Common values of $\beta = 1/9$ (corresponding to $\delta_0/R = 1/10$ for a half-full tumbler, $\phi=0.5$) and $\ell=1/5$ are used in all plots. The values for $\Da$ were selected to best illustrate the features of the model. The most obvious feature of the non-Fickian model [Eq.~\eqref{eq:cd_exact_soln_pulse}] is that, for finite $\Da$, the initial pulse takes a nontrivial amount of time to relax, exhibiting persistent discontinuities at $Z=\pm \ell$. For $T=\mathcal{O}(1)$ (i.e., ``long'' times), the non-Fickian and Fickian solutions agree, but the time it takes for them to do so increases with $\Da$. This result (also shown to hold for the IVP on $Z\in(-\infty,+\infty)$ in \cite{rs13}) justifies using a Fickian diffusion model at long times. However, the non-Fickian aspects have non-trivial consequences at early times (here, in comparison to the axial diffusion time) in the transport process.

\section{Discussion}


\paragraph{Analogy to viscoelasticity.}
Consider the following ``constitutive law'' between the axial flux $q_\mathrm{d}$ and concentration gradient $\partial c_\mathrm{d}/\partial z$ of the diffusing particle species:
\begin{equation}
\left[ 1 + \frac{1}{k(1+\beta)} \frac{\partial}{\partial t}\right]q_\mathrm{d} = -D\frac{\beta}{1+\beta}\left[1 + \frac{1}{k\beta} \frac{\partial}{\partial t}\right]\frac{\partial c_\mathrm{d}}{\partial z},
\label{eq:oldroyd}
\end{equation}
Then, Eq.~\eqref{eq:cd_single_eq} can be obtained by combining the latter with the axial conservation of mass (continuity) equation ${\partial c_\mathrm{d}}/{\partial t} + {\partial q_\mathrm{d}}/{\partial z} = 0$. The significance here is that Eq.~\eqref{eq:oldroyd} is a common constitutive relation, termed the \emph{Jeffreys model} \cite{J32} (see also \cite[\S5.2b]{Bird}, wherein a common sign convention for stress and flux is used), relating the stress to the rate of strain in polymeric fluids. Interestingly, in Eq.~\eqref{eq:oldroyd}, the  ``kinematic viscosity'' $D\beta/(1+\beta)$, the ``relaxation time'' $1/[k(1+\beta)]$, and the ``retardation time'' $1/(k\beta)$ are \emph{not} independent (as they would, in principle, be for a polymeric fluid), since they are all functions of $\beta$.

From this analogy to viscoelasticity, an exact solution of the IBVP for Eq.~\eqref{eq:cd_single_eq_nd1} on $z\in[0,+\infty)$ with $c_\mathrm{d}(0,t) = H(t)$ can be immediately obtained by adapting any of the three representations known from non-Newtonian fluid mechanics (see, e.g., \cite{CJ09,c10}); similarly for the IBVP on $z\in[0,1]$  \cite{c13}. These IBVPs can be interpreted as various industrial feed problems. A time-dependent BC, $c_\mathrm{d}(0,t) = c_{\mathrm{d},0}(t)H(t)$, can be accommodated via the convolution theorem for the Laplace transform.  

\paragraph{Connection to fractional (anomalous) diffusion.}

A salient feature of the analogy to a Jeffreys-type viscoelastic model is that the exchange of particles between the bulk (non-diffusing) and surface (diffusing) species can clearly be interpreted as ``memory'' in the axial diffusion process, and quantified through $\beta$ and $k$ given explicitly in Eq.~\eqref{eq:beta} and Eq.~\eqref{eq:k}, respectively.

Specifically, in this analogy to linear viscoelasticity, Eq.~\eqref{eq:oldroyd} can be re-expressed as a flux--gradient relation in the form of a \emph{memory integral}:%
\begin{equation}
q_\mathrm{d}(z,t) = -\int_{0}^t \mathfrak{K}(t-t') \frac{\partial c_\mathrm{d}}{\partial z}(z,t') \,\rd t',
\label{eq:q_int_del_c}
\end{equation}
with the effective kernel 
\begin{equation}
\mathfrak{K}(t-t') = D \left[ 2\delta(t-t') - k\re^{-k(1+\beta)(t-t')} \right].
\label{eq:mem_kernel}
\end{equation}
For the sake of argument, we have assumed that the flux $q_\mathrm{d}$ does not depend on the history of $\partial c_\mathrm{d}/\partial z$ prior to the initiation of flow ($t=0$) to set the lower limit of integration in Eq.~\eqref{eq:q_int_del_c}. Observe that the kernel in Eq.~\eqref{eq:mem_kernel} contains a Maxwell-type (exponential) ``fading memory'' in addition to the time-local contribution (Dirac-delta term) (see \cite[\S5.2b]{Bird} for further details, including discussion of the factor of 2). For $\mathfrak{K}(t-t') = 2D\delta(t-t')$ (no memory), Eq.~\eqref{eq:q_int_del_c} reduces to Fick's first law \cite{Fick1855,BSL}, $q_\mathrm{d} = - D {\partial c_\mathrm{d}}/{\partial z}$, and $D$ is the usual diffusivity. For a granular flow, the constitutive relation between $q_\mathrm{d}$ and $\partial c_\mathrm{d}/\partial z$ is unknown but Fick's first law is often invoked \cite{l54,hchf66,dgkb91}. A memory-integral constitutive equation for axial granular diffusion, as in Eq.~\eqref{eq:q_int_del_c}, is reasonable because granular force networks \cite{jnb96} form and carry ``information'' across the flowing material. The latter can be interpreted as nonlocal action, leading to (possibly) highly-correlated particle distribution statistics (i.e., ``memory'') in the flow \cite{fc07,mbl08}.

Now, what if $\mathfrak{K}$ decayed ``slower than exponentially'' \cite{g48b}? Then, a generic form for $\mathfrak{K}$, which decays as $t-t'\to\infty$, is a power law: 
\begin{equation}
\mathfrak{K}(t-t') = - D_\alpha \frac{1-\alpha}{\Gamma(\alpha)} (t-t')^{\alpha-2},
\label{eq:plaw_kernel}
\end{equation}
where $\Gamma(\cdot)$ is the Gamma function, $\alpha\in(0,1)$ is a real number, and the proportionality constant $D_\alpha (>0)$ does \emph{not} have the meaning of diffusivity. Substituting $\mathfrak{K}$ from Eq.~\eqref{eq:plaw_kernel} into Eq.~\eqref{eq:q_int_del_c}, we arrive at
\begin{equation}
q_\mathrm{d} = - D_\alpha \, \prescript{}{0}{\mathfrak{D}}_t^{1-\alpha} \frac{\partial c_\mathrm{d}}{\partial z}
\label{eq:q_fracder_del_c}
\end{equation}
instead of Eq.~\eqref{eq:oldroyd}, where 
\begin{equation}
\begin{aligned}
\prescript{}{0}{\mathfrak{D}}_t^{1-\alpha} \Xi(\cdots, t) &:= \frac{1}{\Gamma(\alpha)} \frac{\partial}{\partial t} \int_{0}^t \frac{\Xi(\cdots,t')}{(t-t')^{1-\alpha}} \,\rd t'\\
	&\phantom{:}= -\frac{1-\alpha}{\Gamma(\alpha)} \int_{0}^t \frac{\Xi(\cdots,t')}{(t-t')^{2-\alpha}} \,\rd t'
\end{aligned}
\label{eq:rl_int}
\end{equation}
is the \emph{Riemann--Liouville fractional derivative} of order $1-\alpha$ \cite{p99,mk00}, assuming suitable behavior of $\Xi$ as $t'\to t$ to eliminate boundary terms upon differentiating under the integral sign. Substituting Eq.~\eqref{eq:q_fracder_del_c} into the continuity equation, we obtain a \emph{fractional} diffusion equation%
\begin{equation}
\frac{\partial  c_\mathrm{d}}{\partial t} = D_\alpha \, \prescript{}{0}{\mathfrak{D}}_t^{1-\alpha} \frac{\partial^2 c_\mathrm{d}}{\partial z^2}.
\label{eq:tfrac_de}
\end{equation}
Unlike Eq.~\eqref{eq:cd_single_eq}, Eq.~\eqref{eq:tfrac_de} has the attractive property of possessing self-similar solutions in terms of Fox's $H$-function in the similarity variable $x\big/t^{\alpha/2}$ \cite{mk00}. Note that Eq.~\eqref{eq:tfrac_de} is not exactly in the form of the anomalous axial granular diffusion equation posed in \cite{km05} due to technicalities regarding initial conditions for fractional PDEs. However, we interpret the intent in \cite{km05} to be the same.

\begin{figure}
\centering
\includegraphics[width=\columnwidth]{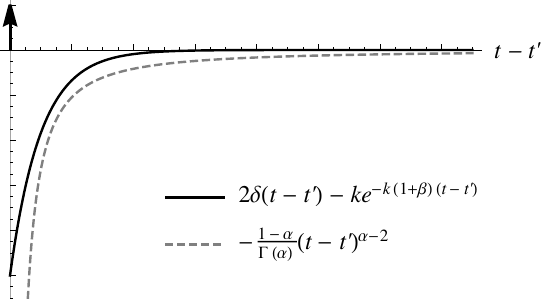}
\caption{Schematic comparison of the shapes of the memory kernels $\mathfrak{K}/D$ from Eq.~\eqref{eq:mem_kernel} (black, solid) and $\mathfrak{K}/D_\alpha$ from Eq.~\eqref{eq:plaw_kernel} (gray, dashed) with $k\simeq k(1+\beta)=10$ \si{\per\second} and $\alpha=2/3$ per \cite{km05}. The vertical arrow at $t-t'=0$  represents the Dirac-delta term in Eq.~\eqref{eq:mem_kernel}.}
\label{fig:kernels}
\end{figure}

Through the connection to the memory integral in Eq.~\eqref{eq:q_int_del_c}, we see that the proposed model [Eq.~\eqref{eq:cd_single_eq}] and the fractional diffusion model [Eq.~\eqref{eq:tfrac_de}] are two cases of a common general theory. A schematic comparison of the shapes of the kernels from Eq.~\eqref{eq:mem_kernel} and Eq.~\eqref{eq:plaw_kernel} is shown in Fig.~\ref{fig:kernels}. Ultimately, only experimental measurements of the kernel $\mathfrak{K}$ would yield a definitive model for axial diffusion (accounting for various ``anomalies'' heretofore discussed). Such measurements remain a challenging open problem in granular mechanics.

\paragraph{Further refinements.}

Finally, in this work, we focused on transport phenomena in a partially-filled cylindrical tumbler (fill fraction $\phi\lesssim 0.5$), which has received the most attention in the experimental studies discussed in Sec.~\ref{sec:intro}. It is well known that, if $\phi>0.5$, then a ``core'' of quasi-static granular material forms along the cylinder axis \cite{metcalfe95,g01,socie05}. Further, a non-circular tumbler cross-section \cite{hkgmo99,col10,kit16} also significantly modifies the mixing and segregation dynamics. In principle, a large fill fraction or a non-circular cross-section would modify the rate constant $k$, and perhaps necessitate unequal exchange rates in each of Eqs.~\eqref{eq:react_diff}. While these details do not fundamentally change the proposed flowing layer--fixed bed exchange mechanism, it would be of interest to address these details in future work.

\section*{Acknowledgement}
This paper is dedicated to memory of the late Prof.~Denis L.\ Blackmore for his contributions to continuum models of granular flows (e.g., \cite{bsr99,rosato16}). I.C.C.\ would also like to acknowledge discussions (circa 2014) on wave phenomena \cite{rs13} with the late Prof.~A.~M.~Samsonov.

This work was initiated while I.C.C.\ was working with H.A.S.\ at Princeton University. At that time, I.C.C. was supported by the U.S.\ National Science Foundation (NSF) under grant no.~DMS-1104047. I.C.C.\ would like to further acknowledge the hospitality of the University of Nicosia, Cyprus, where this version was completed thanks to a Fulbright U.S.\ Scholar award from the U.S.\ Department of State.


\section*{Research data availability}
No data was generated as part of this study.

\printcredits


\bibliographystyle{cas-model2-names}


\balance


\end{document}